\documentclass[prl,fleqn,twocolumn,showpacs,citeautoscript,superscriptaddress]{revtex4-1}

\usepackage{amsmath}    
\usepackage{hyperref}
\usepackage[varg]{txfonts}   
\usepackage{graphicx}   

\usepackage{dcolumn}
\newcolumntype{.}{D{.}{.}{-1}}
\newcolumntype{d}[1]{D{.}{.}{#1}}

\newcommand{\mum}{\mbox{\ensuremath{\mu}m}}
\newcommand{\rnm}{nm$^{-1}$}
\newcommand{\etal}{\textit{et al.}}
\let\mc\multicolumn

\begin{document}

\preprint{\today}

\title{Plasmons in Sodium under Pressure: Increasing Departure from Nearly-Free-Electron Behavior}

\author{I. Loa}
\email[Corresponding author:~E-mail~]{I.Loa@ed.ac.uk}
\altaffiliation[Present address:~]{SUPA, School of Physics and Astronomy,
Centre for Science at Extreme Conditions, The University of Edinburgh, United
Kingdom.}

\author{K. Syassen}
\affiliation{Max-Planck-Institut f\"{u}r Festk\"{o}rperforschung, Heisenbergstr.\ 1,
70569 Stuttgart, Germany}

\author{G. Monaco}

\author{G. Vank\'{o}}
\altaffiliation[Present address:~]{KFKI Research Institute for Particle and
Nuclear Physics, PO Box 49, H-1525 Budapest, Hungary.}

\author{M. Krisch}
\author{M. Hanf\/land}

\affiliation{European Synchrotron Radiation Facility, BP 220, 38043 Grenoble
Cedex, France}

\date{\today}

\begin{abstract}
We have measured plasmon energies in Na under high pressure up to 43~GPa using
inelastic x-ray scattering (IXS). The momentum-resolved results show clear
deviations, growing with increasing pressure, from the predictions for a
nearly-free electron metal. Plasmon energy calculations based on
first-principles electronic band structures and a quasi-classical plasmon model
allow us to identify a pressure-induced increase in the electron-ion
interaction and associated changes in the electronic band structure as the
origin of these deviations, rather than effects of exchange and correlation.
Additional IXS results obtained for K and Rb are addressed briefly.
\end{abstract}

\pacs{
71.45.Gm, 
62.50.-p, 
78.70.Ck, 
71.20.-b  
}

\maketitle

Sodium, at ambient conditions, is one of the best manifestations of a
``simple'' or nearly-free-electron (NFE) metal \cite{WS34}. It is
characterized by a single $s$-type valence electron, weak interaction
between the conduction electrons and the atomic cores (electron-ion
interaction), and conduction band states of $sp$ orbital character. Na
crystallizes in the high-symmetry body-centered cubic (bcc) structure at
pressures up to 65~GPa, where it transforms to face-centered cubic (fcc)
\cite{HLS02}. The properties of Na change fundamentally under pressure in
the megabar pressure range, where a series of phase transitions into
lower-symmetry crystal structures has been predicted \cite{NA01+CN01} and
observed \cite{HSLC02,GLMG08,LGMG09}, accompanied by marked changes in its
optical properties \cite{HSLC02,LGMG09,LGSC09} and culminating in the
formation of a non-metallic, visually transparent phase at $\sim$200~GPa
\cite{MEOX09}. A central question is how the transformation from a simple
metal to a semiconductor progresses, not only in Na, but also in other
metals such as Li, which was reported to become semiconducting above
70~GPa \cite{MS09}. As for Na, does it remain NFE-like in its bcc and fcc
phases up to $\sim$100~GPa \cite{LGSC09} so that the non-NFE behavior
starts only with the transitions into the lower-symmetry phases above
105~GPa, or are there significant precursors at lower pressure?

To provide an answer, we measured and calculated the pressure dependence
of Na plasmon energies. Plasmon excitations provide information on the
collective electronic excitations in the form of longitudinal charge
density waves at finite wavevector, and they determine the optical
response of a metal, specifically the plasma reflection edge. Plasmons
have been studied for many years by electron energy loss spectroscopy
(EELS) at zero pressure (see for example \cite{FSF89,SFF89} and references
therein), but this technique is not suitable for samples enclosed in
high-pressure cells. Mao \etal\ \cite{MKH01} have demonstrated the
possibility of measuring plasmon excitations in Na under pressure using
inelastic x-ray scattering (IXS), and they found their experimental
results up to 2.7~GPa to be in agreement with theoretical predictions.

We report here detailed IXS results on the plasmon energy dispersion in Na
under pressure up to 43~GPa, corresponding to a 2.6-fold increase in
density. Our results evidence a significant departure from the predictions
for a NFE metal. In order to explain this discrepancy between theory and
experiment, we also present plasmon energy calculations based on
first-principles electronic band structures and a quasi-classical plasmon
model after Paasch and Grigoryan (\emph{PG model}) \cite{PG99}. These
calculations reconcile experiment and theory and allow us to identify
changes in the electron-ion interaction as the dominant effect, rather
than changes in the electron-electron interactions. Some experimental
results are also reported for K and Rb.

The theoretical description of plasmons in simple metals is well
established. The most commonly used approach starts from the free-electron
(FE) gas and uses the Random Phase Approximation (RPA) \cite{Pin64}. The
plasmon energy dispersion $E_p(q)$ is then given by $E_p(q) = \hbar
\omega_p + \frac{\hbar^2}{m}\, \alpha q^2$ with the plasma frequency
$\omega_p = \sqrt{{n e^2}/{\epsilon_0 \epsilon_s m}}$ and the dispersion
coefficient $\alpha_\text{FE} = \tfrac{3}{5}{E_F}/{\hbar\omega_p}$, where
$q$ is the plasmon momentum, $m$ the electron mass, $n$ the electron
density, $\epsilon_s$ a dielectric constant describing the polarizability
of the ionic cores ($\epsilon_s = 1$ for the free-electron gas), and $E_F
= (\hbar^2/2m)(3\pi^2n)^{2/3}$ is the Fermi energy. This relatively simple
model works reasonably well for simple metals such as Na and Al at ambient
conditions \cite{FSF89,SFF89}, but experimental dispersion coefficients
$\alpha$ tend to be lower than the theoretical values, which has been
attributed to electron exchange and correlation effects \cite{SFF89}. Both
the plasma frequency $\omega_p$ and the dispersion coefficient $\alpha$
depend on the electron density and can thus be tuned by the application of
pressure; both are expected to increase with increasing pressure. The
stability of bcc sodium over a large pressure range of 0--65~GPa permits
to generate an up to threefold increase in (electron) density without a
structural transition.

IXS experiments were performed on beamline ID16 at the ESRF, Grenoble.
Silicon crystal monochromators were used to monochromatize the incident
beam and to analyze the scattered radiation. The incident x-ray beam with
a photon energy of 9.877~keV was focussed onto the polycrystalline sample
in a high-pressure cell with a spot diameter of 100--200~\mum, depending
on the sample size. IXS spectra of the samples at room temperature were
recorded  in energy-scanning mode with an overall spectral resolution of
0.6~eV and a momentum resolution of 0.4~\rnm. In most of the experiments,
the samples were pressurized in diamond anvil cells (DACs). Rhenium and
stainless steel gaskets were used with initial thicknesses of
50--100~\mum\ and hole diameters of 150--200~\mum. The loading of
distilled Na, K, and Rb into the pressure cells was carried out in an
argon atmosphere. Because of the softness of these metals no pressure
transmitting medium was added. Pressures were determined with the ruby
method \cite{ruby:method} or by measuring an x-ray powder diffraction of
the sample and using its known equation of state \cite{HLS02}. In the
experiments using a DAC, the incoming x-ray beam passed through one
diamond anvil onto the sample, and the scattered radiation was collected
through the opposing anvil (thickness $\sim$1.5~mm). Despite the small
sample scattering volume and the relatively low diamond transmittance of
30\% for 10-keV x-rays, high-quality spectra could be collected in $\sim$2
hours. The experiments on Na at 1 and 7~GPa were performed using a
high-pressure cell equipped with sintered-diamond anvils and a beryllium
gasket. Here, the x-rays passed through the Be gasket.

\begin{figure}[bt]
   \includegraphics[scale=1]{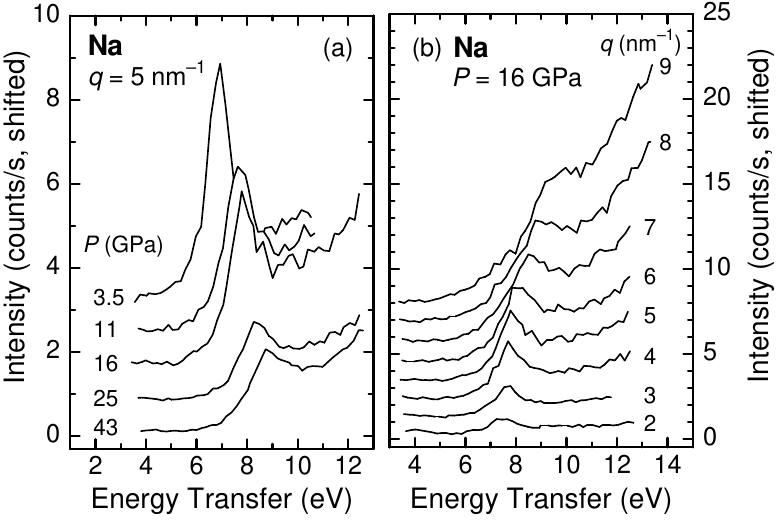}
   \caption{IXS spectra of polycrystalline sodium pressurized in a diamond
   anvil cell. (a) Energy transfer spectra for a momentum transfer of $q =
   5$~\rnm\ and pressures of 3.5--43~GPa. (b) Energy transfer spectra of Na at
   16~GPa and momentum transfers of 2--9~\rnm. Vertical offsets are added for
   clarity.}
   \label{fig:spectra}
\end{figure}

Figure~\ref{fig:spectra} shows IXS spectra of polycrystalline Na under
pressure. A single excitation peak is observed, which is attributed to the bulk
plasmon in Na, based on the predictions for a NFE metal with the electron
density of Na and also by comparison with previous EELS results \cite{FSF89}.
The rising background in the spectra of Fig.~\ref{fig:spectra}, in particular
at large $q$, is due to plasmon and interband excitations in the diamond anvil
through which the scattered radiation is detected \cite{WKAF00}.

\begin{figure}
   \includegraphics[scale=1]{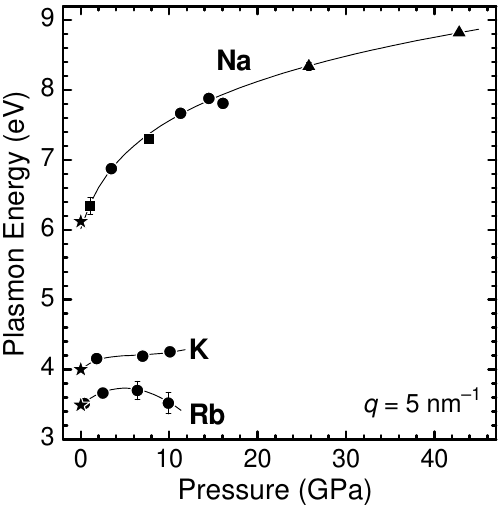}
   \caption{Plasmon energies in Na, K, and Rb at $q = 5$~\rnm\ as a
   function of pressure. Ambient-pressure EELS results \cite{FSF89} are indicated by
   stars. For the Na data, different symbols distinguish
   different sample loadings and pressure cells. Lines are guides to the eye.}
   \label{fig:NaKRb_energies}
\end{figure}

Plasmon energies and linewidths were determined by fitting the spectra with a
Gaussian peak for the plasmon line and a polynomial background. As all
experiments were performed on polycrystalline samples, the reported plasmon
energies are directional averages. The Na plasmon energies increase with
increasing pressure as illustrated in Fig.~\ref{fig:NaKRb_energies} for $q =
5$~\rnm. This is in \emph{qualitative} agreement with the NFE picture, where
the plasmon energies scale with the electron density.

Figure~\ref{fig:NaKRb_energies} also shows results for K and Rb. These two
metals clearly do not follow the expectations for NFE metals. Moreover,
their plasmon linewidths increased and their plasmon intensities decreased
rapidly with increasing pressure. These effects are attributed to the
pressure-driven $s$--$d$ hybridization of conduction band states, as is
also evident from the optical response of K and Rb under pressure
\cite{TTS82,TS83} (see also \cite{KE99}). We will therefore focus on Na
that could be studied over the largest pressure range and which, at
ambient conditions, is one of the best manifestations of a
nearly-free-electron metal.

\begin{figure}
   \includegraphics[scale=1]{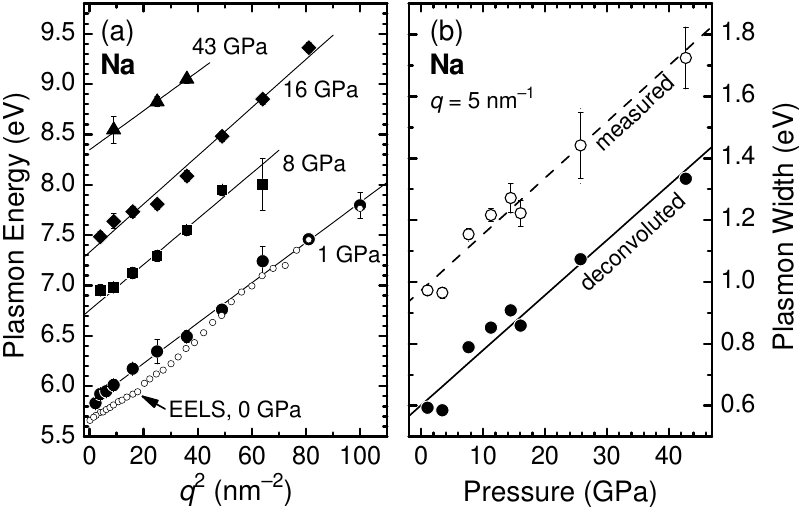}
   \caption{(a) Plasmon dispersions $E(q^2)$ of polycrystalline Na as a
      function of pressure. Ambient-pressure electron-energy-loss spectroscopy
      results by vom Felde \etal\ \cite{FSF89} are indicated by small open symbols. (b) Plasmon
      linewidth of Na versus pressure.
   }
   \label{fig:Na_disp_width}
\end{figure}

Figure~\ref{fig:Na_disp_width}(a) shows the measured plasmon dispersion
relations of Na at several pressures. The plasmon energies are plotted
versus $q^2$ because of the anticipated parabolic dispersion relation, see
the $E_p(q)$ relation given above. At the lowest pressure, 1~GPa, the Na
plasmon dispersion measured here is indeed very close to parabolic.
Results of an ambient-pressure EELS study \cite{FSF89} are included in
Fig.~\ref{fig:Na_disp_width}(a) for comparison. In the low-$q$ region, the
IXS and EELS data are reasonably consistent, but the EELS results exhibit
some deviation from a parabolic dispersion. The offset, at low $q$,
between the IXS and EELS data is largely due to the pressure applied in
the IXS experiment. The \mbox{1-GPa} plasmon dispersion measured by IXS is
described best by $E(0) = \hbar\omega_p = 5.82(2)$~eV and $\alpha =
0.26(1)$. As noted before \cite{FSF89}, the experimental dispersion
parameter $\alpha$ is lower than the FE/RPA value of $\alpha_\text{RPA} =
0.35$. Towards higher pressures, the measured plasmon dispersion remains
approximately parabolic, deviations being most notable at 16~GPa.

Figure~\ref{fig:Na_disp_width}(b) shows that also the plasmon linewidth is
strongly pressure dependent. This effect is tentatively attributed to a
reduction of the plasmon lifetime due to decays involving electron-hole
excitations. A detailed study by, e.g., time-dependent density functional
theory could be a subject for future studies. For the remainder of this paper
we focus on the plasmon energies, assuming that self-energy effects on the
plasmon frequency can be neglected.

\begin{figure}
   \includegraphics[scale=1]{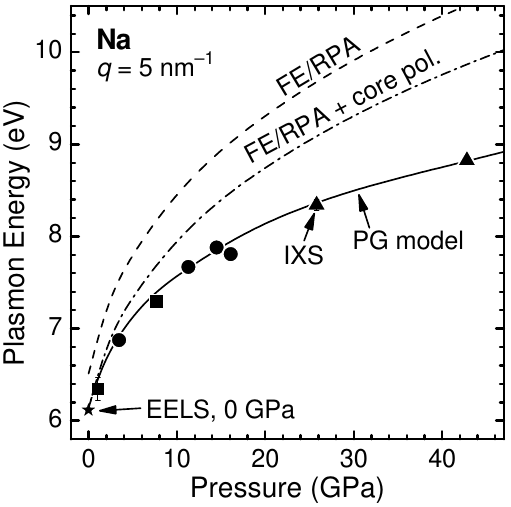}
   \caption{Experimental pressure dependence of the directionally-averaged plasmon
      energy in Na at $q = 5$~\rnm\ (large symbols) and results of the
      free-electron (FE) gas model, FE model with core polarization, and the PG
      model (solid line). The star indicates the ambient-pressure EELS result
      \cite{FSF89}.}
   \label{fig:Na5nm_p-dep}
\end{figure}

The IXS results on the plasmon energies of Na as a function of momentum
and pressure allow us to test the validity of the FE/RPA description of Na
and to assess the relative importance of band structure effects, core
polarizability, and exchange-correlation. Using $E_p(q)$ as given above
and the experimental equation of state of Na \cite{HLS02}, the pressure
dependence of the $q=5$~\rnm\ plasmon was calculated within the FE/RPA
framework as shown in Fig.~\ref{fig:Na5nm_p-dep}. This FE/RPA estimate is
significantly higher in energy than the experimental values, and the
deviation increases with increasing pressure. Inclusion of the core
polarization ($\epsilon_s = 1.16$, Ref.~\cite{NP82}) leads to a good
agreement with the experiment at low pressure, but a major deviation
between theory and experiment remains at high pressure.

The important observation here is the striking increase in the deviation
between theory and experiment with increasing pressure. In previous work on
other metals, deviations from the FE/RPA predictions were discussed in relation
to exchange and correlation effects \cite{FSF89,SGBB91}, and a number of
extensions of the RPA  were proposed in this spirit (see
\cite{SGBB91,FSF89,VS72} and references therein). Their main effect is to
reduce the plasmon dispersion coefficient $\alpha$, and their inclusion can
improve the agreement between theory and experiment. However,
exchange-correlation effects \emph{decrease} with increasing electron density,
and they can thus be excluded as the origin of the deviation between the
`FE/RPA + core polarization' results and the experimental data in
Fig.~\ref{fig:Na5nm_p-dep}. More recent theoretical studies have emphasized the
importance of band-structure effects on the plasmon properties, regarding both
the energy dispersion \cite{PG99} and the plasmon linewidth \cite{KE99}. The
present IXS results offer an opportunity to test these proposals.

The PG plasmon model \cite{PG99} adopted here is an extension of the
FE/RPA approach. This classical model is not expected to describe the
plasmon properties as accurately as, e.g., time-dependent density
functional theory (DFT) \cite{KE99}, but it allows us better to understand
the underlying physics. The electronic structure of the metal is described
here by a single isotropic conduction band with a quartic dispersion,
$E(k) = E_2 k^2 + E_4 k^4$. The $E_4 k^4$ term accounts for deviations
from the parabolic band shape of the free-electron gas. The plasma
frequency $\omega_p$ and the plasmon dispersion coefficient $\alpha$ can
then be determined from $E_2$, $E_4$ and the Fermi energy $E_F$ as
described in detail in \cite{PG99}.

Electronic structure calculations of Na were performed in the framework of
first-principles DFT, using the full-potential \mbox{L/APW+lo} method
\cite{SNS00,MBSS01} and the Generalized Gradient Approximation
\cite{PBE96} as implemented in the WIEN2K code \cite{soft:Wien2k}. The
electronic band structure of bcc Na was calculated for a series of volumes
corresponding to the pressure range of 0--50~GPa
\cite{misc:CompDetails}\nocite{PBE96}. The coefficients $E_2$ and $E_4$
were determined for three directions in the Brillouin zone, i.e.\ along
$[\xi00]$, $[\xi\xi0]$, and $[\xi\xi\xi]$, and then averaged
\cite{SupplMat}. The only adjustable parameter in the calculation of the
plasma frequency and the dispersion coefficient is the static dielectric
constant $\epsilon_s$ that accounts for the polarizability of the ionic
cores. A value of $\epsilon_s = 1.16$ as determined for Na in a DFT-based
study \cite{NP82} was adopted here, and a possible density dependence was
neglected.

\begin{table}[b]
   \caption{Experimental plasma frequencies $\omega_p$ and dispersion coefficients $\alpha$
   of Na as a function of pressure and comparison with theoretical results.
   The experimental values were determined from linear fits to the $E(q^2)$ data as shown
   in Fig.~\ref{fig:Na_disp_width}(a).}
   \label{tab:plasparams}
   \begin{tabular}{D{.}{.}{2.1}@{\extracolsep{1.1em}}D{.}{.}{1.5}D{.}{.}{1.5}@{\extracolsep{1.1em}}D{.}{.}{1.2}D{.}{.}{1.2}@{\extracolsep{1.1em}}D{.}{.}{1.2}D{.}{.}{1.2}}
   \hline
   \hline
         & \mc{2}{c}{experiment}\ & \mc{2}{c}{FE/RPA} & \mc{2}{c}{PG model}    \\
         \cline{2-3} \cline{4-5} \cline{6-7}
   \mc{1}{c}{$P$}    & \mc{1}{c}{$\hbar \omega_p$} & \mc{1}{c}{$\alpha$} & \mc{1}{c}{$\hbar \omega_p$} & \mc{1}{c}{$\alpha$} & \mc{1}{c}{$\hbar\omega_p$} & \mc{1}{c}{$\alpha$}      \\
   \mc{1}{c}{(GPa)}  & \mc{1}{c}{(eV)}   &          & \mc{1}{c}{(eV)}             &          & \mc{1}{c}{(eV)}
&               \\
   \hline
   1.0  & 5.82(2) & 0.26(1) & 5.84 & 0.35 & 5.82 & 0.31 \\
   7.7  & 6.8(1)  & 0.26(2) & 6.99 & 0.37 & 6.77 & 0.31 \\
   16.1 & 7.3(1)  & 0.32(2) & 7.74 & 0.39 & 7.35 & 0.29 \\
   42.8 & 8.3(1)  & 0.26(2) & 9.10 & 0.41 & 8.33 & 0.26 \\
   \hline
   \hline
   \end{tabular}
\end{table}

Figure~\ref{fig:Na5nm_p-dep} shows the results of the PG model with input
from our band structure calculations for the $q=5$~\rnm\ plasmon in Na
\cite{SupplMat}. The computed results are in excellent agreement with the
IXS data. To trace the source of the difference between the FE/RPA results
and those of the PG model, Table~\ref{tab:plasparams} summarizes the
plasma frequency $\hbar \omega_p$ and the dispersion coefficient $\alpha$
for the two models at selected pressures. At 1~GPa, the results of the
NFE/RPA and the PG model are very similar, and the calculated plasma
frequency agrees well with the experiment. As noted before \cite{FSF89},
the calculated values of $\alpha$ are 20--30\% larger than in the
experiment, and this is probably due to exchange-correlation effects not
included here. The effect of the non-parabolic contribution ($E_4 < 0$) is
to reduce both $\hbar \omega_p$ and $\alpha$. In Na, this effect is very
small near ambient pressure, confirming the analysis by Paasch and
Grigoryan \cite{PG99}. With increasing pressure, however, $|E_4|$
increases and leads to substantial corrections \cite{SupplMat}. At 40~GPa,
it reduces the plasma frequency by 8\% and the dispersion coefficient by
35\% compared to NFE/RPA. Table~\ref{tab:plasparams} also shows that
$\alpha$ \emph{decreases} with increasing pressure in the PG model, in
contrast to the free-electron behavior. We would like to emphasize that it
is the non-parabolic contribution that causes the renormalization of the
plasmon energies, even though the corrections to the band energies are
less than 3\% of the band width \cite{SupplMat}.

As for the physical origin of these corrections, the band structure
calculations show that pressure causes the band gaps at the N, P, and H
points of the Brillouin zone to grow relative to the width of the
conduction band \cite{SupplMat}.\ This evidences a strengthening of the
interaction between the valence electrons and the ionic cores
(electron-ion interaction). In other words, Na becomes increasingly less
free-electron like under compression. The quartic correction ($E_4<0$)
corresponds to a lowering of the conduction band energies near the
Brillouin zone boundary in comparison to the free-electron case, and this
distortion of the band structure \cite{SupplMat} leads to a reduction of
the Fermi velocity, $v_F = (1/\hbar)\partial E(k)/\partial k|_{k_F} =
(2k_F/\hbar)(E_2+2E_4k_F^2)$, which is the key physical quantity that
determines the plasmon dispersion.

In summary, we have performed inelastic x-ray scattering experiments to
determine the effect of pressure on the plasmon excitations in the `simple
metal' sodium. While Na is considered one of the best manifestations of a
nearly-free-electron metal at ambient conditions, our results evidence
substantial and increasing deviations from the behavior of a NFE metal at
high pressure up to 43~GPa. This can be seen as an early precursor of the
fundamental changes in the electronic structure of Na at megabar
pressures. The deviation from NFE behavior can in part be attributed to
the polarizability of the ionic cores, but to a larger extent it is caused
by pressure-induced changes in the electronic band structure. Plasmon
energies determined on the basis of electronic band structure calculations
and the quasi-classical PG model are in excellent agreement with the IXS
results. They show that the electron-ion interaction in Na increases with
pressure and leads to the renormalization of the plasmon energies via a
modification of the electronic band structure. As for the heavy alkali
metals under pressure, band structure effects can be expected to be even
more important due to the pressure-driven hybridization of the valence $s$
orbitals with $d$ states, as discussed before \cite{PG99}. We observed a
weak pressure dependence of the plasmon frequencies of K and Rb combined
with fast broadening of their plasmon resonances under pressure. These
results may aid the interpretation of plasmon dispersions of the heavy
alkali metals at ambient pressure. As for bulk Li metal, the quite
detailed predictions on the collective electronic response under pressure
\cite{RSBE08,ERRS10} still await a related experimental investigation.

We thank F.~K\"{o}gel (MPI-FKF, Stuttgart) for providing the distilled metals
used in this study. G.~V. was supported by the Hungarian Scientific
Research Fund (contract No.\ K72597).


\begin{thebibliography}{10}

\bibitem{WS34} E. Wigner and F. Seitz, Phys. Rev. {\bf 46},  509  (1934).

\bibitem{HLS02} M. Hanfland, I. Loa, and K. Syassen, Phys. Rev. B {\bf
    65},  184109  (2002).

\bibitem{NA01+CN01} J.~B. Neaton and N.~W. Ashcroft, Phys. Rev. Lett. {\bf
    86}, 2830 (2001); N.~E.
  Christensen and D.~L. Novikov, Solid State Commun. {\bf 119}, 477 (2001).

\bibitem{HSLC02} M. Hanfland {\it et~al.}, \textit{Sodium at megabar
    pressures}, {Poster} at
  2002 High Pressure Gordon Conference.

\bibitem{GLMG08} E. Gregoryanz {\it et~al.}, Science {\bf 320},  1054
    (2008).

\bibitem{LGMG09} L.~F. Lundegaard {\it et~al.}, Phys. Rev. B {\bf 79},
    064105  (2009).

\bibitem{LGSC09} A. Lazicki {\it et~al.}, Proc. Nat. Acad. Sci. USA {\bf
    106},  6525  (2009).

\bibitem{MEOX09} Y. Ma {\it et~al.}, Nature {\bf 458},  182  (2009).

\bibitem{MS09} T. Matsuoka and K. Shimizu, Nature {\bf 458},  186  (2009).

\bibitem{FSF89} A. vom Felde, J. Spr\"{o}sser-Prou, and J. Fink, Phys.
    Rev. B {\bf 40},  10181
  (1989).

\bibitem{SFF89} J. Spr\"{o}sser-Prou, A. vom Felde, and J. Fink, Phys.
    Rev. B {\bf 40},  5799
  (1989).

\bibitem{MKH01} H.-K. Mao, C. Kao, and R.~J. Hemley, J. Phys.: Condens.
    Matter {\bf 13},  7847
  (2001).

\bibitem{PG99} G. Paasch and V.~G. Grigoryan, Ukr. J. Phys. {\bf 44},
    1480  (1999).

\bibitem{Pin64} D. Pines, {\em Elementary excitations in solids} (W.A.
    Benjamin, New York,
  1964).

\bibitem{ruby:method} G.~J. Piermarini, S. Block, J.~D. Barnett, and R.~A.
    Forman, J. Appl. Phys.
  {\bf 46}, 2774 (1975); H.~K. Mao, J. Xu, and P.~M. Bell, J. Geophys. Res.
  {\bf 91}, 4673 (1986).

\bibitem{WKAF00} S. Waidmann {\it et~al.}, Phys. Rev. B {\bf 61},  10149
    (2000).

\bibitem{TTS82} H. Tups, K. Takemura, and K. Syassen, Phys. Rev. Lett.
    {\bf 49},  1776  (1982).

\bibitem{TS83} K. Takemura and K. Syassen, Phys. Rev. B {\bf 28},  1193
    (1983).

\bibitem{KE99} W. Ku and A.~G. Eguiluz, Phys. Rev. Lett. {\bf 82},  2350
    (1999).

\bibitem{NP82} R. Nieminen and M. Puska, Physica Scripta {\bf 25},  952
    (1982).

\bibitem{SGBB91} L. Serra {\it et~al.}, Phys. Rev. B {\bf 44},  1492
    (1991).

\bibitem{VS72} P. Vashishta and K.~S. Singwi, Phys. Rev. B {\bf 6},  875
    (1972).

\bibitem{SNS00} E. Sj\"{o}sted, L. Nordstr\"{o}m, and D.~J. Singh, Solid
    State Commun. {\bf
  114},  15  (2000).

\bibitem{MBSS01} G.~K.~H. Madsen {\it et~al.}, Phys. Rev. B {\bf 64},
    195134  (2001).

\bibitem{PBE96} J.~P. Perdew, K. Burke, and M. Ernzerhof, Phys. Rev. Lett.
    {\bf 77},  3865
  (1996).

\bibitem{soft:Wien2k} P. Blaha {\it et~al.}, {\em {\bf WIEN2k}, An
    Augmented Plane Wave + Local
  Orbitals Program for Calculating Crystal Properties} (K. Schwarz, Techn.
  Universit\"at Wien, Austria, 2001).

\bibitem{misc:CompDetails} Computational details: sphere size $R_\text{MT}
    = 2.0$~a.u.; plane-wave cutoff
  defined by $R_\text{MT} \times \text{max}(k_n)$ = 8.0; Brillouin zone sampled
  on a tetrahedral mesh with $18^3$ $k$-points (190 in the IBZ).

\bibitem{SupplMat} See EPAPS Document No.\ ***** for additional details.
    For more information on
  EPAPS, see http://www.aip.org/pubservs/epaps.html.

\bibitem{RSBE08} A. Rodriguez-Prieto, V.~M. Silkin, A. Bergara, and P.~M.
    Echenique, New J.
  Phys. {\bf 10},  053035  (2008).

\bibitem{ERRS10} I. Errea {\it et~al.}, Phys. Rev. B {\bf 81},  205105
    (2010).

\end{thebibliography}


\end{document}